\documentstyle[aps,epsfig,multicol]{revtex}

\begin{document} 
 
\title{Exact solution and magnetic properties of an anisotropic spin ladder} 
\author{Zu-Jian Ying$^{1,2,3}$, Itzhak Roditi$^1$, Angela Foerster$^3$, 
 Bin Chen$^2$} 
\address{1. Centro Brasileiro de Pesquisas F\'\i sicas, Rua Dr. Xavier  
Sigaud 150, 22290-180 Rio de Janeiro, RJ, Brasil\\ 
2. Hangzhou Teachers College, Hangzhou 310012, China\\ 
3. Instituto de F\'\i sica da UFRGS, Av. Bento Gon\c calves, 9500, Porto 
Alegre, 91501-970, Brasil\\} 

\date{\ } 

\maketitle 
 
\begin{abstract} 
We study an integrable two-leg spin-1/2 ladder with an XYZ-type rung 
interaction. Exact rung states and rung energies are obtained for the 
anisotropic rung coupling in the presence of a magnetic field. Magnetic 
properties are analyzed at both zero and finite temperatures via the 
thermodynamic Bethe ansatz and the high-temperature expansion. According  
to different couplings in the anisotropic rung interaction,  
there are two cases in which a gap opens, with the ground state  
involving one or two components in the absence of a magnetic field.  
We obtain the analytic expressions of all critical fields for the  
field-induced quantum phase transitions (QPT). 
Anisotropic rung interaction leads to such effects as  
separated magnetizations and susceptibilities in different directions,  
lowered inflection points and remnant weak variation of the magnetization  
after the last QPT. 
\end{abstract} 
 
%\pacs{75.10.Jm, 75.30.Kz, 75.40.Cx} 
 
\begin{multicols}{2} 
 
\section{Introduction} 
 
Recently the quasi-one-dimensional spin ladder has attracted much interest 
both experimentally and theoretically\cite{Dagotto}. More and more 
ladder-structure compounds have been realized, such as  
SrCu$_2$O$_3$\cite{Azuma},  
Cu$_2$(C$_5$H$_{12}$N$_2$)$_2$Cl$_4$\cite{Chaboussant},  
(5IAP)$_2$CuBr$_4\cdot $2H$_2$O\cite{Landee},  
(C$_5$H$_{12}$N)$_2$CuBr$_4$\cite{Watson}, and so forth. Although many  
ladder compounds can be well described by simple isotropic ladders, the  
structural distortion and the spin-orbital interaction of the transition  
ions can lead to various magnetic anisotropies. Besides the spin-orbital  
interaction, both on-site Coulomb exchange  
interaction\cite{on-siteU,on-siteU-Stein} and nonlocal Coulomb 
interaction\cite{neighborU} also influence the anisotropy. Anisotropic 
interaction from bond buckling has been recently found in copper-oxide  
ladder compounds CaCu$_2$O$_3$\cite{CaCuO} due to an angle deviation  
from 180$^{\circ }$ in the Cu-O-Cu bond\cite{CaCuO,Yushankhai,Kataev}.  
An anisotropic rung interaction was considered in Ref.\cite{Citro}  
motivated by CaCu$_2$O$_3$\cite{CaCuO} and a two-leg spin ladder with  
an XXZ-rung interaction was derived in the presence of the  
Dzyaloshinskii-Moriya interaction and the  
Kaplan-Shekhtman-Entin-Wohlman-Aharony interactions. When the Cu-O-Cu 
bond is near 90$^{\circ }$, the rung interaction is weak in the  
copper-oxide ladder. Spin anisotropy in the exchange interaction  
also exists in strongly-coupled ladder compounds such as  
(pipdH)$_2$CuBr$_4$\cite{Patyal}. On the other hand, real spin ladder  
compounds are usually described by the standard Heisenberg ladder model, 
which is not exactly soluble, turning the computation of the physical  
properties for the ground state (GS), the gap, the thermodynamical  
quantities and other relevant properties in the presence of temperature  
and magnetic fields, rather difficult. Usually, just numerical  
calculations and perturbative schemes can be applied. Recently it was  
shown that the integrable spin ladder model\cite{WangY} can describe  
the properties of strongly-coupled spin ladder compounds when a rescaling  
parameter is introduced\cite{XiwenPRL,XiwenSU4}. Therefore, it can be  
expected that integrable ladders with anisotropic rung interactions  
can provide some meaningful information in the physics of anisotropies.  
In addition, anisotropic rung interactions also provide us with some  
more adjustable parameters that may be useful in fitting the 
experimental data of compounds. 
 
In the present paper we shall consider the anisotropy in the rung 
interaction and the corresponding magnetic anisotropic effect by solving an 
integrable spin ladder with a general XYZ rung interaction. By means of the 
thermodynamical Bethe ansatz (TBA) and the high temperature expansion  
(HTE)\cite{XiwenPRL,Tsuboi,Shiroishi}, we investigate the influence of the 
anisotropic rung interaction on the quantum phase transitions (QPT) and 
the magnetic properties. The contents are arranged as follows: (i) In the 
second section we present the model and the exact rung-state basis in the 
presence of a magnetic field. Then the model is solved by 
the Bethe ansatz (BA) approach. (ii) The third section gives the TBA 
equations for the GS and the HTE for physical properties at finite  
temperatures. (iii) In the forth section, we study the field-induced  
QPT's. The rung anisotropy provides two kinds of gapped ladders,  
respectively with one and two components in the GS. The 
analytic expressions are obtained for all the critical fields of the 
corresponding QPT's. The rung anisotropy also leads to a separation of  
the magnetizations and susceptibilities in different directions. The  
magnetization inflection point (IP) may be lowered from the  
half-saturation and in the two-component gapped ladder the IP is  
even not invariant under different temperatures. A remnant variation  
of magnetization can be found after the last QPT. In the fifth section  
we give a summary of our results. 
 
\section{The model, exact rung states and BA solution} 
 
We shall consider a spin-1/2 two-leg spin ladder model with a general 
XYZ-type anisotropy in the rung interaction, whose Hamiltonian reads  
\begin{eqnarray} 
{\cal H} &=&{\cal H}_0+{\cal H}_{XYZ}+{\cal M},\   \nonumber  \\ 
{\cal H}_0 &=&\frac{J_0}\gamma \sum_{i=1}^L P_{i,i+1},   \nonumber \\ 
{\cal H}_{XYZ} &=&\sum_i(J_xS_i^xT_i^x+J_yS_i^yT_i^y+J_zS_i^zT_i^z),  
\nonumber \\ 
{\cal M} &=&-gH\sum_i(S_i^z+T_i^z),  \label{H} 
\end{eqnarray} 
where $\vec S$ and $\vec T$ are the spin operators for the two legs and $g$ 
is the Land\'e $g$ factor in the direction of the field. $J_0$ is the 
average of the leg interaction and $\gamma $ is a rescaling parameter.  
The value $\gamma =4$ was introduced in Refs.\cite{XiwenSU4,XiwenPRL}  
in fitting with isotropic ladder compounds in the presence of weak rung.  
The bulk part ${\cal H}_0$ with the permutation operator  
$P_{i,i+1}=(2\vec S_i\cdot \vec S_{i+1}+\frac 12)(2\vec T_i\cdot  
\vec T_{i+1}+\frac 12)$ exhibits the SU(4) symmetry\cite{LiSU4}.  
Isotropic integrable spin ladder\cite{WangY} has identical  
$J_x$, $J_y$ and $J_z$. 
 
When the rung interaction is strong, it is favorable for the spin ladder 
system to form rung states, since the leg interaction is too weak to take 
apart the rung state. Anisotropy in the rung interaction leads to the 
collapse of the conventional singlet and triplet rung states from the 
isotropic ladder, even in the absence of the field. However, we find a new 
exact basis valid both in the absence and presence of an external magnetic 
field,  
\begin{eqnarray} 
\varphi _1 &=&\frac 1{\sqrt{2}}\left( \left| \uparrow \downarrow 
\right\rangle -\left| \downarrow \uparrow \right\rangle \right) ,\quad 
\varphi _2=\frac 1{\sqrt{2}}\left( \left| \uparrow \downarrow \right\rangle 
+\left| \downarrow \uparrow \right\rangle \right)   \nonumber \\ 
\varphi _3 &=&\frac{\left| \uparrow \uparrow \right\rangle -\eta ^{-1}\left| 
\downarrow \downarrow \right\rangle }{\sqrt{1+\eta ^{-2}}},\quad \varphi _4= 
\frac{\left| \uparrow \uparrow \right\rangle +\eta \left| \downarrow 
\downarrow \right\rangle }{\sqrt{1+\eta ^2}} 
\end{eqnarray} 
where  
\begin{equation} 
\eta ^{\pm 1}=\frac{\pm 4gH+\sqrt{(4gH)^2+(J_x-J_y)^2}}{J_x-J_y}. 
\label{eta} 
\end{equation} 
Then the corresponding rung energies include the Zeeman energy in a  
nonlinear way,  
\begin{eqnarray} 
E_1 &=&-\frac 14(J_x+J_y+J_z),  \nonumber \\ 
E_2 &=&\frac 14(J_x+J_y-J_z),  \nonumber \\ 
E_3 &=&-\sqrt{(gH)^2+\frac 1{16}(J_x-J_y)^2}+\frac 14J_z,  \nonumber \\ 
E_4 &=&\sqrt{(gH)^2+\frac 1{16}(J_x-J_y)^2}+\frac 14J_z. 
\end{eqnarray} 
 
The rung states \{$\varphi _i\mid i=1,\cdots ,4$\} provides a new 
fundamental representation of the SU(4) Lie algebra $S_m^n\varphi _i=\delta 
_{n,i}\varphi _m$ with commutation relations of the generators  
$[S_m^n,S_k^l]=\delta _{n,k}S_m^l-\delta _{m,l}S_k^n$. Based on this SU(4) 
realization and the vanishing commutation relations  
$[{\cal H}_{XYZ},{\cal H}_0]=[{\cal M},{\cal H}_0]=0$, one can solve the  
model (\ref{H}) via the BA approach\cite{Sutherland}. The BA equations are  
the same as those obtained for the SU(4) model\cite{LiSU4BA} and  
for the SU(3)$\otimes $U(1) spin ladder\cite{WangY}. Here we present  
the BA equations together with the eigenenergy  
\begin{eqnarray} 
-\prod_{m=1}^{M^{(k)}}\Xi _1(\mu _{j,m}^{k,k}) &=&\prod_{m=1}^{M^{(k+1)}}\Xi 
_{\frac 12}(\mu _{j,m}^{k,k+1})\prod_{m=1}^{M^{(k-1)}}\Xi _{\frac 12}(\mu 
_{j,m}^{k,k-1}),  \nonumber \\ 
E &=&-\sum_{j=1}^{M^{(1)}}2\pi a_1(\mu _j^{(1)})+\sum_{i=1}^4E_iN_i, 
\label{BAE} 
\end{eqnarray} 
where $\Xi _x(\mu _{j,m}^{k,l})=(\mu _j^{(k)}-\mu _m^{(l)}-xi)/(\mu 
_j^{(k)}-\mu _m^{(l)}+xi)$, $\mu _j^{(0)}=0,M^{(0)}=L,M^{(4)}=0,$ and $1\leq 
k\leq 3;$ $a_n(\mu )=\frac 1{2\pi }\frac n{\mu ^2+n^2/4}$. There are $L$ 
rungs, $N_i$ is the total number of rung state $\varphi _i$ and  
$\mu _j^{(k)}$ is the rapidity. $M^{(k)}$ is the total rapidity number  
in the $k$'th branch. 
 
From the BA equations and the eigenenergy we can apply the TBA and HTE to 
study the collective properties of the model. When the field is applied in  
the $x$ or $y$ direction, one only needs to permute the anisotropy parameter 
values $\left\{ J_x,J_y,J_z\right\} $ as well as the corresponding $g$ 
factors. We incorporate $g$ into the field unit in plotting figures to 
investigate the net effect of the anisotropic rung interaction. For a powder 
sample, a simple way is to take an average of the three directions. 
 
\section{TBA and HTE} 
 
By adopting the string conjectures\cite{string} and applying the Yang-Yang  
method\cite{Y-Y} at the thermodynamic limit, one can obtain the GS  
equations for three dressed energies $\epsilon ^{(i)}$,  
\begin{equation} 
\epsilon ^{(i)}=g^{(i)}-a_2*\epsilon ^{(i)-}+a_1*(\epsilon 
^{(i-1)-}+\epsilon ^{(i+1)-}),  \label{TBA} 
\end{equation} 
where $\epsilon ^{(0)}=\epsilon ^{(4)}=0$ and the symbol $*$ denotes the 
convolution. The basis order is chosen as ($\varphi _{P_1}\varphi 
_{P_2}\varphi _{P_3}\varphi _{P_4}$)$^T$, where $P_i\in \{1,2,3,4\}$ and  
$\varphi _{P_1}$ is energetically the most favorable state while  
$\varphi _{P_4}$ is the least favorable one. For the chosen order the  
driving term is given by $g^{(i)}=E_{P_{i+1}}-E_{P_i}$. The GS is composed  
of Fermi seas filled by negative $\epsilon ^{(i)-}$. If some branch of the  
dressed energy is all positive, then the corresponding excitations to this  
branch is gapful. A QPT occurs at the point where the gap is closed. We  
shall apply these TBA equations to analyze the field-induced QTP for the GS. 
 
For the finite temperature case the TBA involves an infinite number of coupled  
integral equations. In the present paper we shall apply the  
HTE\cite{Shiroishi,Tsuboi,XiwenPRL} from T-system\cite{Kuniba} within the  
Quantum Transfer Matrix formalism\cite{QTM}, which involves only a finite  
number of integral equations and consequently is more convenient. Following  
Refs.\cite{XiwenPRL,Shiroishi,Tsuboi}, one can obtain the free energy  
$f$ for per rung in high temperatures. Here we present the first four terms  
which dominate the physics for high temperatures:   
\begin{equation} 
f=-T\left( C_0+C_1(\frac{J_0}{\gamma T})+C_2(\frac{J_0}{\gamma T})^2+C_3 
(\frac{J_0}{\gamma T})^3\right)   \label{freeE} 
\end{equation} 
where $T$ is the temperature, the coefficients are  
\begin{eqnarray*} 
C_0 &=&\ln Q_{+},\ C_1=\frac{2Q}{Q_{+}^2},\ C_2=\frac{3Q}{Q_{+}^2}-
\frac{6Q^2}{Q_{+}^4}+\frac{3Q_{-}}{Q_{+}^3}, \\ 
C_3 &=&\frac{10Q}{3Q_{+}^2}-\frac{18Q^2}{Q_{+}^4}+\frac{80Q^3}{3Q_{+}^6}+ 
\frac{8Q_{-}}{Q_{+}^3}-\frac{24QQ_{-}}{Q_{+}^5}+\frac 4{Q_{+}^4}, 
\end{eqnarray*} 
with the definitions 
\begin{eqnarray*} 
Q &=&2\cosh (\frac 12\beta J_z)+4\cosh (\frac 14\beta J_{x+y})\cosh (\beta 
h), \\ 
Q_{\pm } &=&2e^{(\pm \beta J_z/4)}\cosh (\frac 14\beta J_{x+y})+2e^{(\mp 
\beta J_z/4)}\cosh (\beta h), 
\end{eqnarray*} 
$h=\sqrt{(gH)^2+J_{x-y}{}^2/16}$, $J_{x\pm y}=J_x\pm J_y$ and $\beta =1/T$. 
One can get higher orders for lower temperatures.  
The magnetization and the susceptibility can be easily 
obtained by $M=-\partial f/\partial H,\ \chi =\partial M/\partial H$. The 
rescaling parameter $\gamma =4$ in the isotropic case\cite{XiwenSU4,XiwenPRL} 
gives the leading term of the gap for the integrable ladder in fitting with 
the experimental ladder compounds. If the rung $J_0$ is weak, then the HTE  
gives a valid result even for low temperatures due to the large rescaling  
$\gamma $.  
 
\section{Phase transitions and magnetic properties} 
 
\subsection{One-component gapped ladder} 
 
\begin{figure}[p] 
\setlength\epsfxsize{75mm}
\epsfbox{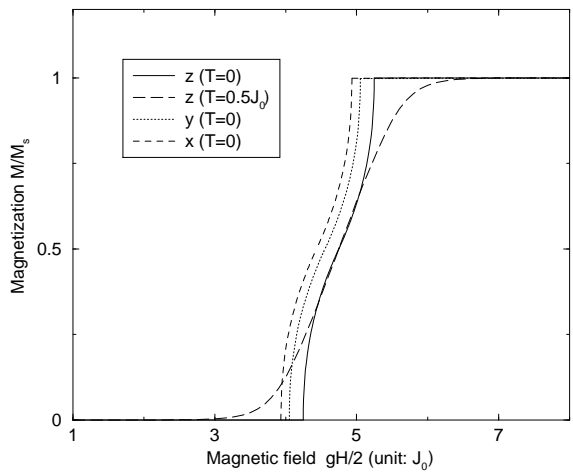}
%%%%%%\resizebox{0.4\textwidth}{!}{\includegraphics{Mz8910.eps} } \vspace*{0cm} 
\caption{Magnetization versus the magnetic field at zero temperature for the  
one-component gapped ladder with a weakly-anisotropic rung interaction,  
$J_x=8J_0$, $J_y=9J_0$, $J_z=10.5J_0$, with the rescaling $\gamma =4$.  
$M_s$ is the saturation magnetization. In the gapped phase $H<H_{c1}$,  
only one component $\varphi _1 $ exists in the ground state. To study the  
net effect of the anisotropic rung interaction, we incorporate the $g$ factor  
into the field. The weak anisotropy in the rung separates the magnetization in  
different directions. The zero temperature magnetization is obtained from the  
thermodynamical Bethe ansatz (TBA). Also for comparison with the finite  
temperature case, a magnetization at $T=0.5J_0$ obtained from the  
high-temperature expansion (HTE) is presented in the $z$ direction. } 
\label{Mz8910} 
\end{figure}

\begin{figure}[p] 
\setlength\epsfxsize{75mm}
\epsfbox{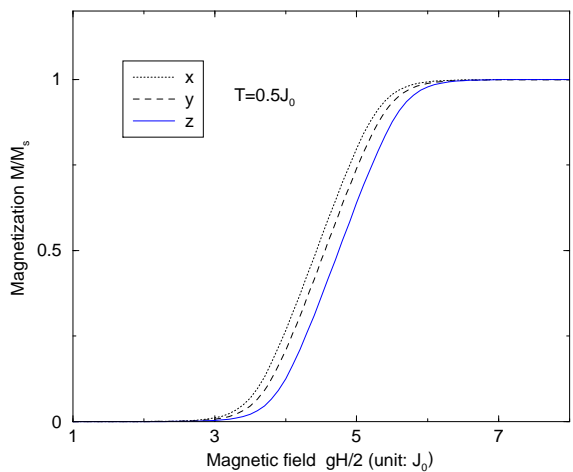}
%%%\resizebox{0.4\textwidth}{!}{\includegraphics{MzT8910.eps} } \vspace*{0cm} 
\caption{Magnetization versus the magnetic field for  $T=0.5J_0$ for the  
one-component gapped ladder with the weak anisotropy, $J_x=8J_0$, $J_y=9J_0$,  
$J_z=10.5J_0$ and $\gamma =4$. The magnetizations are obtained from the HTE,  
which coincides with the magnetization separation in the zero temperature  
case obtained from the TBA. } 
\label{MzT8910} 
\end{figure}

\begin{figure}[p] 
\setlength\epsfxsize{75mm}
\epsfbox{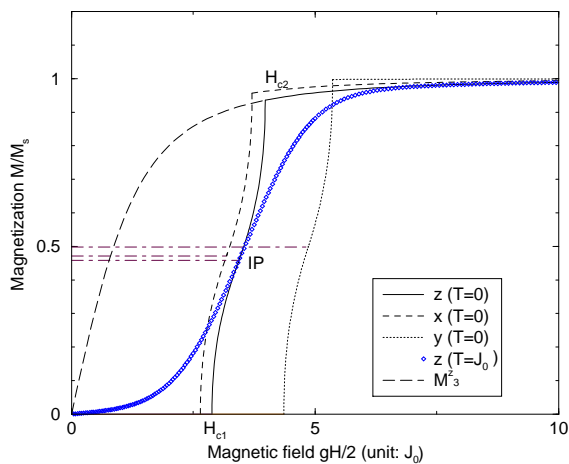}
%%%\resizebox{0.4\textwidth}{!}{\includegraphics{Mz3615.eps} } \vspace*{0cm} 
\caption{Magnetization versus the magnetic field at zero temperature for  
the one-component gapped ladder with a strongly-anisotropic rung interaction,  
$J_x=3J_0$, $J_y=15J_0$, $J_z=6J_0$, $\gamma =4$. The strong anisotropy leads  
to a strong separation of the magnetization. The inflection point (IP) is  
lowered from the half saturation. A temperature magnetization from HTE is  
present to demonstrate the IP. Note that the magnetizations do not reach the  
saturation after the second quantum phase transition at $H_{c2}$, there still  
remains a weak variation of magnetizations. This remnant magnetization  
variation comes from the single-state magnetization $M_3^z$ of $\varphi _3$  
which is a mixture of full-polarized states  
$\left| \uparrow \uparrow \right\rangle $ and the 
lowest-magnetized state $\left| \downarrow \downarrow \right\rangle $. The 
variation of $M_3^z$ is illustrated by the long-dashed line for the whole 
process of the field application. } 
\label{Mz3615} 
\end{figure}

\begin{figure}[p] 
\setlength\epsfxsize{75mm}
\epsfbox{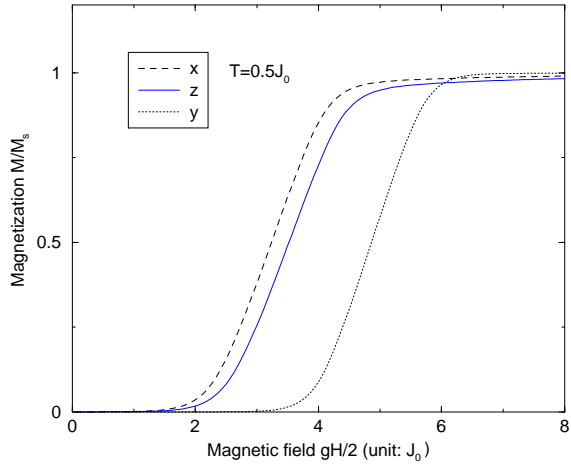}
%%%\resizebox{0.4\textwidth}{!}{\includegraphics{MzT13615.eps} } \vspace*{0cm} 
\caption{ 
Magnetizations versus the magnetic field for different directions in  
temperature case for the one-component gapped ladder with the strong  
anisotropy, $J_x=3J_0$, $J_y=15J_0$, $J_z=6J_0$, $\gamma =4$. } 
\label{MzT13615} 
\end{figure}

For different anisotropies there are two different gapped ladders. In 
one case, only one component $\varphi _1$ exists in 
the gapped GS when $H=0$. In the other case both $\varphi _1$ and $\varphi _2 
$ are involved in the gapped GS. First we discuss the former case which 
happens more likely. It requires  
\begin{eqnarray} 
J_z+J_3-\left| J_x-J_y\right|  &>&16J_0/\gamma ,  \label{gapped1} \\ 
J_x+J_y &>&8J_0/\gamma ,  \label{gapped2} 
\end{eqnarray} 
where $J_3=J_z+J_y+J_z$. Condition (\ref{gapped1}) expels the components  
$\varphi _3$ and $\varphi _4$ from the GS, while condition (\ref{gapped2}) 
excludes the component $\varphi _2$. The field will bring $\varphi _3$ down 
to the ground state and close the gap  
$\Delta =\min \{E_2,E_3,E_4\}-E_1-4J_0/\gamma $ at a 
critical field $H_{c1}$, which leads to the first quantum phase transition 
(QPT). Further increase of the field will bring all components of $\varphi _1$ 
out of the GS and another gap  
$\Delta =E_1-E_3-4J_0/\gamma $ opens at the critical field $H_{c2}$, which  
characterizes another QPT. The factor $4J_0/\gamma $ in the 
gap comes from the maximum depth of the first dressed energy branch. It is 
easy to see that only the components $\varphi _1$ and $\varphi _3$ compete in 
the GS involving one branch of dressed energy, since the GS only consists of  
$\varphi _1$ in the absence of the field while only $\varphi _3$ is lowered 
in energy level when the field is applied. The analytic expressions of two 
critical fields can be obtained exactly as  
\begin{eqnarray} 
H_{c1} &=&\frac 1{2g}\sqrt{J_zJ_3+J_xJ_y+64(\frac{J_0}\gamma )^2-8\frac{J_0} 
\gamma \left( J_z+J_3\right) },  \nonumber \\ 
H_{c2} &=&\frac 1{2g}\sqrt{J_zJ_3+J_xJ_y+64(\frac{J_0}\gamma )^2+8\frac{J_0} 
\gamma \left( J_z+J_3\right) }.  \label{Hc2} 
\end{eqnarray} 
Setting $J_z=J_y=J_z $ recovers the result for isotropic case\cite{XiwenSU4}, 
as expected. 
A weak anisotropy will lead to different critical fields and consequently 
separate the magnetizations in different directions. We give an example of 
the magnetization with weak anisotropy in Fig.\ref{Mz8910}, a 
low-temperature magnetization was presented for comparison in the $z$ 
direction. The corresponding low-temperature magnetizations for all three 
directions are presented in Fig.\ref{MzT8910}, obtained from the HTE. 
Magnetizations in different directions for strong anisotropy are 
demonstrated as an example in Fig.\ref{Mz3615} for the GS and in  
Fig.\ref{MzT13615} for a low temperature. 
 
\begin{figure}[p] 
\setlength\epsfxsize{75mm}
\epsfbox{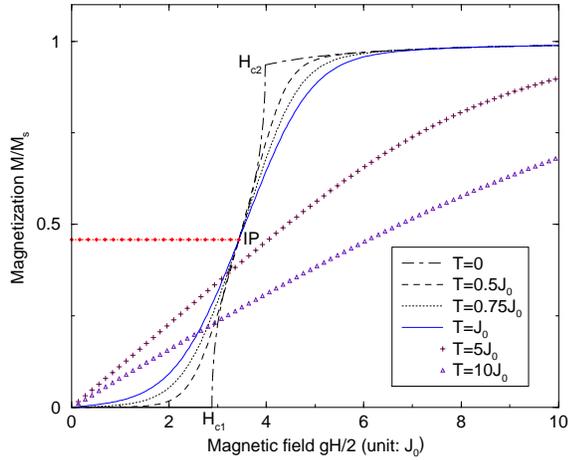}
%%%\resizebox{0.4\textwidth}{!}{\includegraphics{MzT3615.eps} } \vspace*{0cm} 
\caption{ 
Magnetization versus the magnetic field at different temperatures for the 
strongly-anisotropic rung, $J_x=3J_0$, $J_y=15J_0$, $J_z=6J_0$, $\gamma =4$. 
Low temperature magnetizations in $T=0$, $0.5J_0$, $0.75J_0$ and $J_0$ cross 
the inflection point (IP). Higher-temperature magnetizations at $T=5J_0$ and  
$10J_0$ do not go through the IP, as the gap for excitation to $\varphi _2$ 
is overcome by the temperature stimulation.  
} 
\label{MzT3615} 
\end{figure}

Before the gap is closed at $H_{c1}$, the gap $\Delta $ near $H_{c1}$ can be 
expanded to a simpler form 
\begin{equation} 
\Delta \cong c_1\left( H_{c1}-H\right), 
\end{equation} 
$c_1=g^2H_{c1}/\sqrt{(gH_{c1})^2+\frac 1{16}(J_x-J_y)^2}$. Considerable 
excitations can be stimulated by the temperature $T$ if $T$ is in the order 
of the gap $T\sim (H_{c1}-H)$, the magnetization will rise from zero before 
the field reaches the critical point. An expansion based on small Fermi 
points\cite{XiwenSU4} gives the zero-temperature critical behavior in the 
vicinity of $H_{c1}$  
\begin{equation} 
\left\langle M^z\right\rangle \cong \left\langle M^z\right\rangle _3\frac 1 
\pi \sqrt{\frac{c_1}{J_0/\gamma }}(H-H_{c1})^{1/2}. 
\end{equation} 
Here $\left\langle M^z\right\rangle _3$ is the magnetization of a single 
rung state $\varphi _3$, it also varies with the field due to the 
anisotropic rung interaction, as we will discuss below in (\ref{Mz3}). For the 
lowest order in the critical behavior, $\left\langle M^z\right\rangle _3$ 
takes the value at the critical point $H_{c1}$. This $M^z\propto 
(H-H_{c1})^{1/2} $ critical behavior, typical for gapped integer spin 
antiferromagnetic chains\cite{Affleck}, is buried by the afore-mentioned 
temperature effect. This temperature effect can be seen in Fig.\ref{MzT3615},  
in which the magnetization of the $z$ direction at $T=0.5J_0$ 
becomes considerable at the field $H=H_{c1}-0.5J_0$. Actually the magnetization 
at $T=0.5J_0$ increases nearly in a linear way before the $H_{c1} $. 
 
A special point in the magnetization is the inflection point (IP) $H_{IP}$,  
which is an invariant point under low temperatures, 
\begin{equation} 
gH_{IP}=\frac 12\sqrt{(J_z+J_x)(J_z+J_y)}  \label{IP-H} 
\end{equation} 
where the two components $\varphi _1$ and $\varphi _3$ have the same rung 
energies $E_1=E_3$ and the same proportion $N_1=N_3$ in the GS. The excitations  
to $\varphi _2$ and $\varphi _4$ are gapful, the gap can be obtained exactly 
from (\ref{TBA})  
\begin{eqnarray} 
\Delta _{IP} &=&\min \{\Delta _{IP2},\ \ \Delta _{IP4}\},  \nonumber \\ 
\Delta _{IP2} &=&(J_x+J_y)/2-(2\ln 2)J_0/\gamma ,  \nonumber \\ 
\ \Delta _{IP4} &=&(J_z+J_3)/2-(2\ln 2)J_0/\gamma .  \label{gapIP} 
\end{eqnarray} 
At low temperatures, the excitations to $\varphi _2$ or $\varphi _4$  
are difficult 
to stimulate, while the temperature does not influence the relative 
proportion between $\varphi _1$ and $\varphi _3$ due to their same energies at 
the IP. Consequently the proportions of $\varphi _1$ and $\varphi _3$ remain 
almost unchanged when the temperature varies. Therefore the magnetization at  
$H_{IP}$ also remains unmoved when the temperatures changes, and the 
magnetization curves of various temperatures cross each other at the same 
point $M_{IP}$, which is shown by the curves for temperatures $T=0$,  
$0.5J_0$, $0.75J_0$ and $J_0$ in Fig.\ref{MzT3615}. This requires low  
temperatures  
\begin{equation} 
T\ll \Delta _{IP}  \label{T-IP} 
\end{equation} 
as well as the gapped ladder conditions (\ref{gapped1}) and (\ref{gapped2}),  
$\Delta _{IP}$ is the excitation gap to $\varphi _2$ or $\varphi _4$ in (\ref 
{gapIP}). When the temperature is sufficiently high such that the excitations  
to $\varphi _2$ or $\varphi _4$ are considerable, the involvement of these 
components reduces the proportion of $\varphi _3$ which has the highest 
magnetization. The components $\varphi _2$ and $\varphi _4$ have zero and 
negative magnetizations, respectively. As a result, the magnetization at  
$H_{IP}$ deviates from $M_{IP}$ and move downwards. We show this moving 
by magnetization curves at temperatures $T=10J_0$, $20J_0$ in Fig.\ref 
{MzT3615}, for which one can find observation examples in  
Cu$_2$(C$_5$H$_{12}$N$_2$)$_2$Cl$_4$\cite{Chaboussant}. 
 
The magnetization at the IP can be worked out as  
\begin{equation} 
M_{IP}^z=\frac g2\frac{H_{IP}{}^2+H_{IP}H_{IP}^{(+)}} 
{H_{IP}^{(+)2}+H_{IP}H_{IP}^{(+)}}, 
\end{equation} 
where $H_{IP}^{(+)}=\sqrt{H_{IP}^2+(J_x-J_y)^2/(4g)^2}$. For the isotropic 
ladder, $H_{IP}^{(+)}=H_{IP}$ and consequently $M_{IP}^z$ is located at the 
half of the saturation magnetization $M_s^z=g$\cite{XiwenPRL}.  
The anisotropy lowers the magnetization of the IP due to  
$H_{IP}^{(+)}>H_{IP}$, i.e. $M_{IP}^z/M_s^z<1/2$. Physically, the anisotropy 
in $x$,$y$ directions hybridizes the elemental state $\left| \downarrow 
\downarrow \right\rangle $ into $\varphi _3$ so that $\varphi _3$ is not a 
pure fully-polarized elemental state $\left| \uparrow \uparrow \right\rangle  
$ as in the isotropic case. For XXZ-type rung interaction  
$M_{IP}^z$ is half-saturation when the field is oriented in $z$ direction, 
but also lowered when the field is applied in other directions. This IP 
lowering effect is more obvious for the strong anisotropic case, we give an 
example in Fig.\ref{Mz3615}. As one can see from this figure, besides the 
strong separation of the magnetization in different directions, the IP 
points in $x$ and $z$ directions move below the half saturation point. 
 
In addition to the separation of the magnetization in different directions and 
the lowering of the inflection points, another property in the anisotropic case 
is the remnant variation of the magnetization after the second phase transition. 
The fact that magnetization increases 
between $H_{c1}$ and $H_{c2}$ mainly comes from 
the proportional competition between the two state $\varphi _1$ and $\varphi _3$, 
i.e., more rungs are occupied by $\varphi _3$ when the field gets higher. 
The single-rung magnetization in state $\varphi _3$ can be obtained 
explicitly  
\begin{equation} 
\left\langle M^z\right\rangle _3=g\frac{\eta ^2-1}{\eta ^2+1},  \label{Mz3} 
\end{equation} 
where $\eta $ increases with the field from the expression (\ref{eta}). The 
long-dashed line in Fig.\ref{Mz3615} gives an example of $\left\langle 
M^z\right\rangle _3$ in $z$ direction, which increases from zero from the 
beginning of the application of the magnetic field. If $H_{c1}$ is small, 
then the increment of $\left\langle M^z\right\rangle _3$  also  
contributes with an important part to the growth of the magnetization.  
Otherwise for higher $H_{c1}$, the change of  
$\left\langle M^z\right\rangle _3$ contributes less 
to the growth of the total magnetization, since $\left\langle 
M^z\right\rangle _3$ has slowed down in increasing before the first quantum 
phase transition occurs. However, the competition between $\varphi _1$ and  
$\varphi _3$ comes to end after the second phase transition and the 
magnetization is completely $\left\langle M^z\right\rangle _3$. This  
gives a remnant variation of magnetization even after the second phase 
transition, since $\left\langle M^z\right\rangle _3$ is still approaching 
to the saturation limit. This remnant variation of magnetization is 
illustrated for the ground state in Fig.\ref{Mz3615} and can also be seen 
for the temperature case ( Fig.\ref{MzT13615} ). 
 
\begin{figure}[p] 
\setlength\epsfxsize{75mm}
\epsfbox{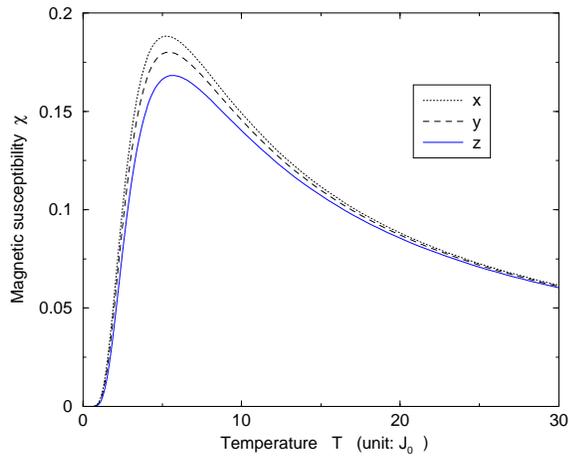}
%%%\resizebox{0.4\textwidth}{!}{\includegraphics{sus8910.eps} } \vspace*{0cm} 
\caption{ 
Magnetic susceptibility against the temperature in different  
directions for the weakly-anisotropic case, $J_x=8J_0$, $J_y=9J_0$,  
$J_z=10.5J_0$ and $\gamma =4 $. The weak anisotropy separate the heights  
of the susceptibility peaks. To see the net effect of the rung anisotropy,  
we plot the figures using the same $g $ factors for the three directions.  
} 
\label{sus8910} 
\end{figure}

\begin{figure}[p] 
\setlength\epsfxsize{75mm}
\epsfbox{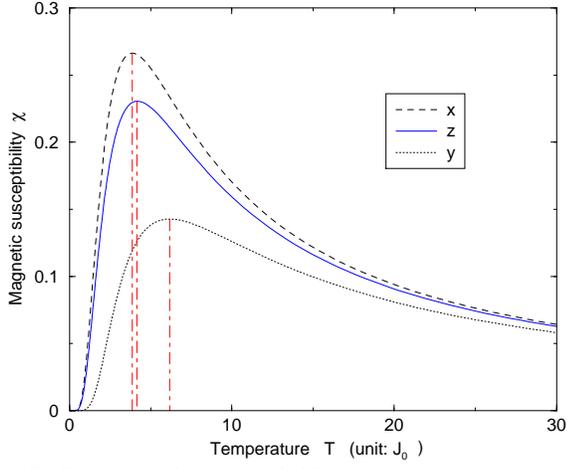}
%%%\resizebox{0.4\textwidth}{!}{\includegraphics{sus3615.eps} } \vspace*{0cm} 
\caption{ 
Magnetic susceptibility versus the temperature in the  
strongly-anisotropic case, $J_x=3J_0$, $J_y=15J_0$, $J_z=6J_0$,  
$\gamma =4$. The strong anisotropy separate not only the peak heights  
but also the whole shape including the peak positions.  
} 
\label{sus3615} 
\end{figure}

Examples of the magnetic susceptibility in the three directions are plotted 
in Fig.\ref{sus8910} for weakly anisotropic rung and in Fig.\ref{sus3615} for 
strongly anisotropic rung. Weak anisotropy separates the heights of the 
magnetic susceptibility peak, while a strong anisotropy also leads to an  
obvious separation of the whole susceptibility including the peak positions. 
 
\subsection{Two-component gapped ladder} 
 
\begin{figure}[p] 
\setlength\epsfxsize{75mm}
\epsfbox{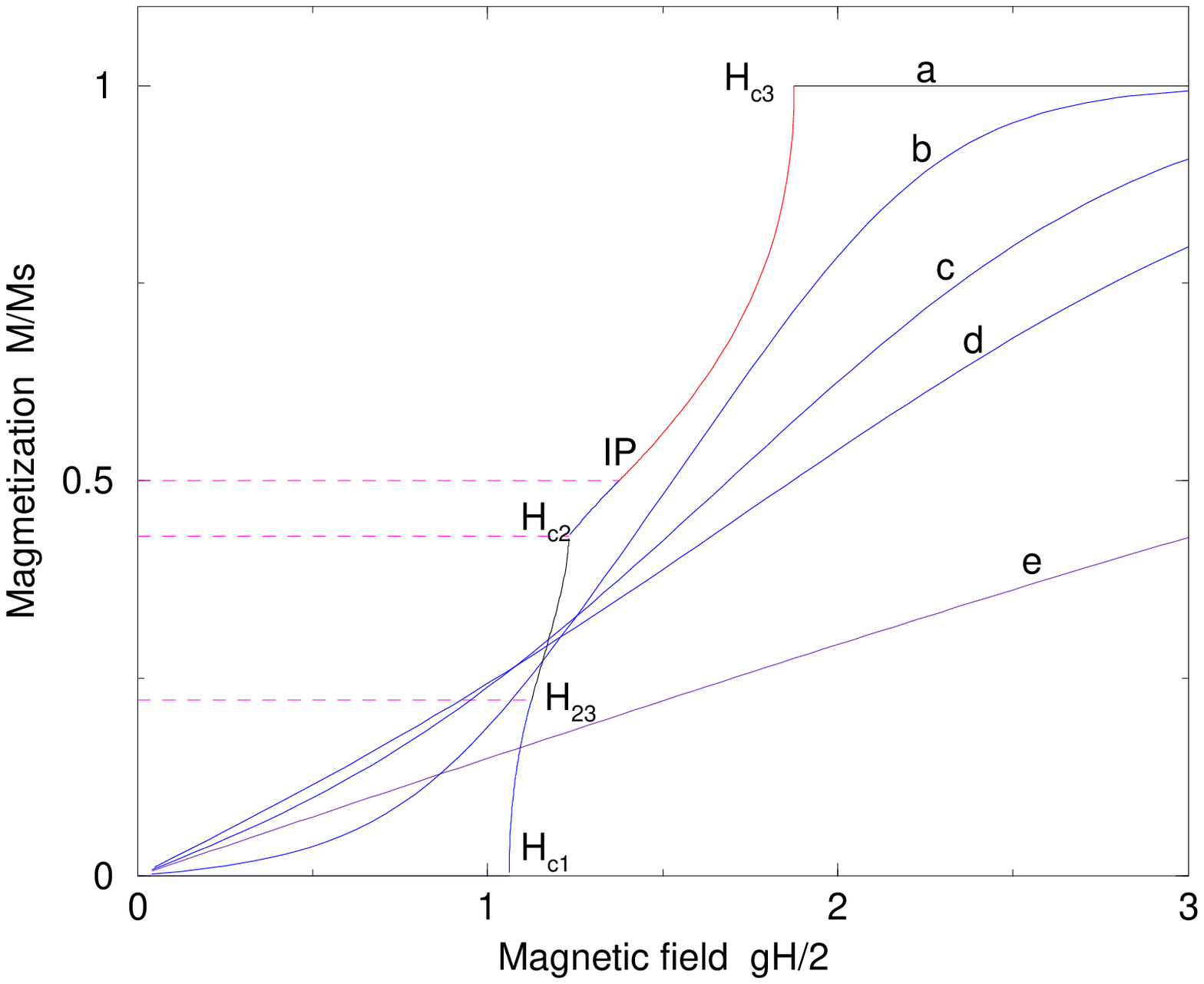}
%%%\resizebox{0.4\textwidth}{!}{\includegraphics{QIP.eps} } \vspace*{0cm} 
\caption{ 
Magnetizations in $z$ direction of a two-component gapped ladder 
case, $J_x=J_y=0.5J_0$, $J_z=5J_0$, $\gamma =4$. The curves {\it a}, {\it b},  
{\it c}, {\it d} and {\it e }are plotted in different temperatures, 
respectively, T$=0$, $0.5J_0${\it , }$J_0$, $1.5J_0${\it \ }and $5J_0${\it .  
}In the gapped phase $H<H_{c1}$, two components $\varphi _1$ and $\varphi _2$ 
involve in the ground state (GS) (T=0). The component $\varphi _3$ begins to 
enter the GS at the critical point $H_{c1}$ and reaches the same energy as  
$\varphi _2$ at $H_{23}$. The component $\varphi _2$ get out of the GS at the 
second critical field $H_{c2}$. The magnetization curves of temperatures do 
not go through the IP as in the one-component gapped ladder case.  When the 
field is applied in $x$ or $y$ direction the GS magnetization will increase 
from the beginning due to the gapless excitation in these directions. 
} 
\label{QIP} 
\end{figure}

Anisotropy in the rung interaction also provides another possibility of a 
gapped ladder, in which not only the rung state $\varphi _1$ but also  
$\varphi _2$ are involved in the GS before the field is applied and brings about 
the first QPT. The single-state energy difference is $E_2-E_1=(J_x+J_y)/2$. 
The larger is the difference, the more strongly $\varphi _1$ and $\varphi _2$ 
will expel each other in the Fermi sea. The two-component gapped ladder 
requires  
\begin{equation} 
\left| J_x+J_y\right| <8J_0/\gamma ,  \label{gapped3} 
\end{equation} 
so that $\varphi _1$ and $\varphi _2$ are close enough in the energy levels to 
exist in the GS at the same time in the absence of the field. Also a strong  
$J_z$ is needed to expel $\varphi _3$ and $\varphi _4$ from the gapped GS 
before the field is applied, approximately  
\begin{equation} 
J_z>4\ln 2\frac{J_0}\gamma +\frac 12\left| J_x-J_y\right| +\frac \gamma  
{8\pi ^2J_0}\left( J_x+J_y\right) ^2. 
\end{equation} 
For simplicity, we assume $J_x+J_y>0$ so that $\varphi _1$ has lower energy 
than $\varphi _2$, one only needs to change $J_x+J_y$ to be $-(J_x+J_y)$ for 
lower $\varphi _2$. The first QPT occurs when the field brings down $\varphi 
_3$ and gets involved in the GS, the critical field can be obtained with 
the help of the Wiener-Hopf technique\cite{W-H} which is valid for large Fermi 
points (Fermi surface in one dimension). Explicitly we have  
\begin{equation} 
gH_{c1}\cong \sqrt{[\frac{J_z}2-2\ln 2\frac{J_0}\gamma -\frac{(J_x+J_y)^2} 
{16\pi ^2J_0/\gamma }]^2-\frac{(J_x-J_y)^2}{16}},  \label{Hc1GS2} 
\end{equation} 
which gives a good approximation if the value of $J_x+J_y$ is not very 
close to $8J_0/\gamma $. Further increase will lower the energy of $\varphi 
_3$ below $\varphi _1$ and $\varphi _2$ and bring them out of the GS one by 
one. The component variations in the QPT are $\{\varphi _1,\varphi 
_2\}\rightarrow \{\varphi _1,\varphi _2,\varphi _3\}\rightarrow \{\varphi 
_1,\varphi _3\}\rightarrow \{\varphi _3\}$, where each arrow indicates the 
occurrence of a QPT. Since $\varphi _2$ has also zero magnetization, the 
total magnetization also remains null in the gapped phase before the first 
QPT. The zero-magnetization component $\varphi _2$ gets out of the GS after  
$H_{IP}$ if  
\begin{equation} 
J_x+J_y<(4\ln 2)J_0/\gamma , 
\end{equation} 
while for  
\begin{equation} 
(4\ln 2)J_0/\gamma <J_x+J_y<8J_0/\gamma ,  \label{C2<IP} 
\end{equation} 
$\varphi _2$ is brought off the GS before $H_{IP}$. These happen at the 
second QPT with an approximate critical field  
\begin{equation} 
gH_{c2}\cong \sqrt{[\frac 12J_z-\frac 34J_{x+y}+4\ln 2\frac{J_0}\gamma + 
\frac{\delta ^2\gamma }{2\pi ^2J_0}]^2-\frac{J_{x-y}^2}{16}},  \label{Hc2GS2} 
\end{equation} 
where $\delta =J_{x+y}-4\ln 2(J_0/\gamma )$. The 
expression (\ref{Hc2GS2}) can give a satisfactory approximation when the 
value of $\left| \delta \right| $ is not near $4J_0/\gamma $. The exact 
critical field $H_{c3}$ for the third QPT is the same as $H_{c2}$ in (\ref 
{Hc2}). When the example in fig.(\ref{QIP}) has numerical points  
$H_{c1}=1.061J_0$ and $H_{c2}=1.232J_0$, the expressions (\ref{Hc1GS2}) and  
(\ref{Hc2GS2}) provide analytic results $H_{c1}=1.064J_0$ and  
$H_{c2}=1.231J_0 $.

The IP in the one-component gapped ladder case will not be invariant in the 
two-component ladder case. If the component $\varphi _2$ gets out of the GS 
after $H_{IP}$, the $\varphi _2$ is gapless. Although the components  
$\varphi _1$ and $\varphi _3$ still have the same proportion at the IP, any 
small temperature will excite more components of $\varphi _2$ and 
consequently decreases the proportion of $\varphi _1$ and $\varphi _3$. 
Therefore the temperature will lower the total magnetization from that of 
the GS. If the component $\varphi _2$ gets out of the GS before $H_{IP}$, 
with the condition (\ref{C2<IP}), the IP is hardly an invariant. 
Despite of the existence of a gap for excitations to $\varphi _2$ at $H_{IP}$,  
the gap is actually quite small  
\begin{equation} 
\Delta _{IP2}<(4-\ln 2)J_0/\gamma , 
\end{equation} 
relative to the strong rung interaction. So a low temperature of order $J_0$  
will still stimulate considerable excitations to $\varphi _2$ and lower the 
magnetization at $H_{IP}$. We illustrate this by an example in  
Fig.(\ref{QIP}).  
 
\section{Summary} 
 
We have introduced a two-leg spin-1/2 ladder with a general anisotropic XYZ 
rung interaction. In particular, the exact rung state basis for this model  
was found. We have studied the effect of the anisotropic rung interaction  
by solving the integrable ladder in the context of the thermodynamical Bethe  
ansatz and the high-temperature expansion. Two kinds of gapped ladders were  
provided, respectively involving one and two components in the GS in the  
absence of the magnetic field. We have obtained analytically all the 
corresponding critical fields for the field-induced quantum phase 
transitions. The magnetizations and susceptibilities in different directions 
separates under the rung anisotropy. The magnetization inflection point is 
lowered from the half-saturation and a weak changing in magnetization still 
remains after the last quantum phase transition. The inflection point in the  
two-component gapped ladder case is not invariant as in the one-component  
gapped ladder case due to field-deduced three-component competition  
or small excitation gap. 
 
\section*{Acknowledgments} 
 
We thank Huan-Qiang Zhou and Xi-Wen Guan for helpful discussions. ZJY 
thanks FAPERJ and FAPERGS for financial support. IR thanks PRONEX and CNPq. 
AF thanks FAPERGS and CNPq. BC is supported by National Nature Science  
Foundation of China under Grant No.10274070 and Zhejiang Natural Science  
Foundation RC02068.

\end{multicols} 
 
%%%%%%%%%%%%%%%%%%%%%%%%%%%%%%%%%%%% Figure 1  
%%%%%%%%%%%%%%%%%%%%%%%%%%%%%%%%% end of Figure 1  
%%%%%%%%%%%%%%%%%%%%%%%%%%%%%%%%%%%% Figure 2  
%%%%%%%%%%%%%%%%%%%%%%%%%%%%%%%%% end of Figure 2 
%%%%%%%%%%%%%%%%%%%%%%%%%%%%%%%%%%%% Figure 3  
%%%%%%%%%%%%%%%%%%%%%%%%%%%%%%%%% end of Figure 3 
%%%%%%%%%%%%%%%%%%%%%%%%%%%%%%%%%%%% Figure 4  
%%%%%%%%%%%%%%%%%%%%%%%%%%%%%%%%% end of Figure 4 
%%%%%%%%%%%%%%%%%%%%%%%%%%%%%%%%%%%% Figure 5  
%%%%%%%%%%%%%%%%%%%%%%%%%%%%%%%%% end of Figure 5 
%%%%%%%%%%%%%%%%%%%%%%%%%%%%%%%%%%%% Figure 6  
%%%%%%%%%%%%%%%%%%%%%%%%%%%%%%%%% end of Figure 6 
%%%%%%%%%%%%%%%%%%%%%%%%%%%%%%%%%%%% Figure 7  
%%%%%%%%%%%%%%%%%%%%%%%%%%%%%%%%% end of Figure 7 
%%%%%%%%%%%%%%%%%%%%%%%%%%%%%%%%%%%% Figure 8  
%%%%%%%%%%%%%%%%%%%%%%%%%%%%%%%%% end of Figure 8 
              

\begin{references} 
\bibitem{Dagotto}  E. Dagotto and T. M. Rice, Science \textbf{271},  
618 (1996); 
E. Dagotto, Rep. Prog. Phys. \textbf{62}, 1525 (1999). 
 
\bibitem{Azuma}  M. Azuma, Z. Hiroi, M. Takano, K. Ishida and Y. Kitaoka, 
Phys. Rev. Lett. \textbf{73}, 3463 (1994). 
 
\bibitem{Chaboussant}  G. Chaboussant, P. A. Crowell, L. P. L\'evy, O. 
Piovesana, A. Madouri, and D. Mailly, Phys. Rev. B \textbf{55},  
3046 (1997). 
 
\bibitem{Landee}  C. P. Landee, M. M. Turnbull, C. Galeriu, J. Giantsidis, 
and F. M. Woodward, Phys. Rev. B \textbf{63}, 100402 (2001). 
 
\bibitem{Watson}  B. C. Watson, V. N. Kotov, M. W. Meisel, D. W. Hall, G. E. 
Granroth, W. T. Montfrooij, S. E. Nagler, D. A Jensen, R. Backov, M. A. 
Petruska, G. E. Fanucci, and D. R. Talham, Phys. Rev. Lett. \textbf{86}, 5168 
(2001). 
 
\bibitem{on-siteU}  T. Yildirim, A. B. Harris, O. Entin-Wohlman, and A. 
Aharony, Phys. Rev. Lett. \textbf{73}, 2919 (1994). 
 
\bibitem{on-siteU-Stein}  J. Stein, O. Entin-Wohlman, and A. Aharony, Phys. 
Rev. B \textbf{53}, 775 (1996). 
 
\bibitem{neighborU}  T. Moriya, Phys. Rev. \textbf{120}, 91 (1960); J. Stein, 
Phys. Rev. B \textbf{53}, 785 (1996). 
 
\bibitem{CaCuO}  V. Kiryukhin, Y. J. Kim, K. J. Thomas, F. C. Chou, R. W. 
Erwin, Q. Huang, M. A. Kastner and R. J. Birgeneau, Phys. Rev. B \textbf{63}, 
144418 (2001). 
 
\bibitem{Yushankhai}  V.Yu. Yushankhai and R. Hayn, Europhys. Lett.  
\textbf{47}, 116 (1999). 
 
\bibitem{Kataev}  V. Kataev, K.-Y. Choi, M. Gr\"uninger, U. Ammerahl, B. 
B\"uchner, A. Freimuth, A. Revcolevschi, Physica B \textbf{312}-\textbf{313},  
614 (2002). 
 
\bibitem{Citro}  R. Citro and E. Orignac, Phys. Rev. B \textbf{65}, 134413 
(2002). 
 
\bibitem{Patyal}  B.R. Patyal, B.L. Scott, and R.D. Willett, Phys. Rev. B  
\textbf{41}, 1657 (1990). 
 
\bibitem{WangY}  Y. Wang, Phys. Rev. B \textbf{60}, 9236 (1999). 
 
\bibitem{XiwenPRL}  M. T. Batchelor, X.-W. Guan, N. Oelkers, K. Sakai, Z. 
Tsuboi, and A. Foerster, Phys. Rev. Lett. \textbf{91}, 217202 (2003). 
 
\bibitem{XiwenSU4}  M.T. Batchelor, X.W. Guan, A. Foerster and H.Q. Zhou, 
New J. Phys. \textbf{5}, 107 (2003). 
 
\bibitem{Tsuboi}  Z. Tsuboi, J. Phys. A \textbf{36}, 1493 (2003). 
 
\bibitem{Shiroishi}  M. Shiroishi and M. Takahashi, Phys. Rev. Lett.  
\textbf{89}, 117201 (2002). 
 
\bibitem{LiSU4}  Y.Q. Li, M. Ma, D.N. Shi and F.C. Zhang, Phys. Rev. Lett.  
\textbf{81}, 3527 (1998). 
 
\bibitem{Sutherland}  B. Sutherland, Phys. Rev. B \textbf{12}, 3795 (1975). 
 
\bibitem{LiSU4BA}  Y.Q. Li, M. Ma, D.N. Shi and F.C. Zhang, Phys. Rev. B  
\textbf{60}, 12781 (1999). 
 
\bibitem{string}  M. Takahashi, Prog. Theor. Phys. \textbf{46}, 401 (1971). 
 
\bibitem{Y-Y}  C.N. Yang and C.P. Yang, J. Math. Phys. \textbf{10},  
1115(1969). 
 
\bibitem{Kuniba}  A. Kuniba, T. Nakanishi, and J. Suzuki, Int. J. Mod. Phys. 
A \textbf{9}, 5215 (1994). 
 
\bibitem{QTM}  M. Suzuki, Phys. Rev. B \textbf{31}, 2957 (1985); A. Kl\"umper, 
Ann. Phys. (Leipzig) \textbf{1}, 540 (1992); G. J\"uttner, A. Kl\"umper, and J. 
Suzuki, Nucl. Phys. B \textbf{487}, 650 (1997). 
 
\bibitem{Affleck}  I. Affleck, Phys. Rev. B \textbf{43}, 3215 (1991). 
 
\bibitem{W-H}  M.G. Krein Usp. Mat. Nauk \textbf{13}, 3 (1958). 
 
\end{references}
\end{document}